\documentclass[prd,twocolumn,showpacs,floatfix,amsmath,amssymb,floatfix]{revtex4}
\usepackage{graphicx,color,dcolumn,booktabs,bm}
\usepackage{longtable,lscape}
\usepackage{txfonts}
\usepackage{overpic}
\usepackage{amssymb}
\usepackage{indentfirst}
\usepackage{feynmf}   
\usepackage{slashed}  
\usepackage{cases}
\usepackage{color}
\usepackage{multirow}
\usepackage{graphicx,color,dcolumn,booktabs,bm}
\usepackage{epsfig,dsfont,amssymb,amsmath,amsfonts,amsbsy,mathrsfs}

\graphicspath{{Figures/}} %

\begin{document}

\title{How can $X^{\pm}(5568)$ escape detection?}

\vspace{1cm}

\author{ Hong-Wei Ke$^1$\footnote{khw020056@hotmail.com}and
        Xue-Qian Li$^2$\footnote{lixq@nankai.edu.cn}  }

\affiliation{  $^{1}$ School of Science, Tianjin University, Tianjin 300072, China \\
  $^{2}$ School of Physics, Nankai University, Tianjin 300071, China }

\vspace{12cm}

\begin{abstract}
Multi-quark states were predicted by Gell-Mann when the quark
model was first formulated. Recently, numerous exotic states that
are considered to be multi-quark states have been experimentally
confirmed (four-quark mesons and five-quark baryons). Theoretical
research indicates that the four-quark state might comprise
molecular and/or tetraquark structures. We consider that the meson
containing four different flavors $su\bar b\bar d$ should exist
and decay via the $X(5568)\to B_s\pi$ channel. However, except for
the D0 collaboration, all other experimental collaborations have
reported negative observations for $X(5568)$ in this golden
portal. This contradiction has stimulated the interest of both
theorists and experimentalists. To address this discrepancy, we
propose that the assumed $X(5568)$ is a mixture of a molecular
state and tetraquark, which contributes destructively to
$X(5568)\to B_s\pi$. The cancellation may be accidental and it
should be incomplete. In this scenario, there should be two
physical states with the same flavor ingredients, with spectra of
$5344\pm307$ and $6318\pm315$. $X(5568)$ lies in the error range
of the first state. We predict the width of the second state
(designated as $S_2$) as $\Gamma(X_{S_2}\to B_s\pi)=224\pm97$ MeV.
We strongly suggest searching for it in future experiments.


\end{abstract}

\maketitle

\section{Introduction}

A resonance named $X(5568)$ was reported by the D0 collaboration
in the $B_s\pi^{\pm}$ invariant mass spectrum, where $B_s$ was
reconstructed by the $J/\psi\,\phi$ final state. The mass and
width were determined as $(5567.8\pm 2.9^{+0.9}_{-1.9})$ MeV and
$(21.9\pm 6.4^{+5.0}_{-2.5} )$ MeV, respectively\cite{D0:2016mwd}.
These new observations have stimulated great interest among both
theorists and experimentalists because analyses indicate that it
should be an exotic state with four different flavors ($ s u \bar
b\bar d$ or its charge conjugates). If this observation is
correct, then it is a non-ambiguous signal of the existence of the
four-quark exotic state, although multi-quark states were
predicted by Gell-Mann when the quark model was first formulated.
Recently, several $X,Y,Z$
particles\cite{Abe:2007jn,Choi:2005,Choi:2007wga,Aubert:2005rm,Ablikim:2013emm,
Ablikim:2013wzq,Ablikim:2013mio,Liu:2013dau,Collaboration:2011gj,Ablikim:2014dxl},
have been discovered. However, in most of these states, the charm
or bottom flavors are hidden, which makes the confirmation of a
four-quark structure difficult. In principle, no law forbids an
exotic state with open charm or bottom flavor. Therefore, we
consider that such a meson (e.g., $X(5568)$) should exist in
nature. In addition, we do not know whether the more favorable
structure is a molecular state or a tetraquark, or even their
mixture. Clearly, only experiments can give us the answer.

Except for the D0 collaboration, all other important experimental
facilities throughout the world, including the LHCb
collaboration\cite{Aaij:2016iev}, the CMS collaboration at Large
Hadron Collider (LHC)\cite{Sirunyan:2017ofq}, and the CDF
collaboration at Fermilab\cite{Aaltonen:2017voc}, have claimed
that no such state can be detected in the $X\to B_s\pi^{\pm}$
channel. At the end of 2017, the D0 collaboration again declared
that they had confirmed the existence of $X(5568)$ from the decay
$X(5568)\to B_s\pi^{\pm}$, where $B_s^0$ was reconstructed via a
semileptonic decay $B_s^0\to
\mu^{\pm}D_s^{\mp}$\cite{Abazov:2017poh}, but its width shifted to
a slightly smaller number $18.6^{+7.9}_{-6.1}({\rm
stat})^{+3.5}_{-3.8}({\rm syst})$ MeV than the value measured
previously. By contrast, the ATLAS
collaboration\cite{Aaboud:2018hgx} very recently announced a
negative observation of $X(5568)$ in the $B_s\pi$ invariant mass
spectrum, i.e., no significant signal was found.

The clear discrepancy between the results obtained by the D0
collaboration and others has led to a dispute because of the
obvious significance of $X(5568)$ for understanding the quark
model, and thus great efforts have made to resolve this issue.
Burns and Swanson suggested\cite{Burns:2016gvy} that an additional
hadron should be undetectable in the $X(5568)$ production process.
In addition, as generally argued, the possibility that $X(5568)$
represents a physical particle comprising four different flavors
cannot be excluded.

In fact, various studies have provide different opinions about the
mysterious $X(5568)$
\cite{Guo:2016nhb,Yang:2016sws,Chen:2016ypj,Albaladejo:2016eps,
Agaev:2016urs,Stancu:2016sfd,Wang:2016mee,Liu:2016ogz,Chen:2016npt,Wang:2016tsi,
Zhang:2017xwc,Kang:2016zmv,Chen:2016mqt,Xiao:2016mho,Agaev:2016ijz,Dias:2016dme,
Wang:2016wkj,Ke:2018jql,Wang:2018jsr,Lang:2016jpk,Lu:2016kxm}. If
it does exist, this clearly raises the question about how it
escapes detection.

It is well known that a four-quark state may be a hadronic
molecule or a tetraquark \cite{Vijande:2009kj}. According to
theoretical computations by several groups, a pure molecular state
or tetraquark makes a substantial contribution to $X(5568)\to
B_s+\pi^{\pm}$ and should be ``seen" by experimental scanning.
Previous studies numerically computed the partial width of the
mode in terms of various phenomenological models by assuming it is
a molecule, whereas others performed computations by assuming that
it is a tetraquark. It is interesting to note that regardless of
whether $X(5568)$ was assumed to be a pure hadronic molecule or a
tetraquark, the numerical estimates of the partial width of
$X(5568)\to B_s\pi$ were remarkably close (and close to the
results obtained by the D0 collaboration). These findings suggest
the following possible scenario. The $s u\bar b\bar d$ exotic
state is a mixture of a molecular state and tetraquark, and they
contribute destructively to the golden channel $X(5568)\to
B_s\pi$. The mixing parameter (mixing angle) determines their
transition amplitudes, which almost cancel each other. We consider
that this cancellation is accidental and incomplete, so a weak
signal should exist. Moreover, the uncertainty is large for the
recently measured width ($X(5568)$ ($18.6^{+7.9}_{-6.1}({\rm
stat})^{+3.5}_{-3.8}({\rm syst})$)), and thus the width may be
relatively narrow. If the width is sufficiently small, the signal
might be drowned in a messy background.


In this study, we admit that four-quark states with $su \bar b\bar
d$ flavors exist in nature and we assume that they are mixtures of
a tetraquark and molecular state with quantum numbers of $0^+$.
This mixing results in two physical states and their quantum
numbers are also $0^+$. We temporally designate one of them as
$X(5568)$. For $X(5568)$, the transition matrix elements of the
two components possess opposite phases, so their contributions to
$X\to B_s\pi$ are destructive. This cancellation leads to a small
partial width that escapes detection. As a consequence, the
existence of another eigenstate with two components that
contribute constructively to the $B_s\pi$ final state would
produce a wide peak. According to our ansatz, the mixing angle of
the two components can be fixed in terms of the theoretical
estimated decay rates for $X(5568)$ given
by\cite{Ke:2018jql,Wang:2016wkj}. By using the bare masses of the
tetraquark\cite{Burns:2016gvy} and molecular
state\cite{Feng:2011zzb}, we obtain the masses of the physical
states by diagonalizing the mass matrix. A rigorous test of this
ansatz involves searching for the partner of $X(5568)$.

In Section II, we discuss how the mixing of a tetraquark and
molecular state produces a physical $X(5568)$ and its partner. In
Section III, we present the numerical results. Finally, we give
our conclusions.

\section{Mixing of a tetraquark and molecular state in a meson}
For a meson that contains four different flavors $s u\bar b\bar
d$, mixing between a molecular state ($M$) and tetraquark ($T$)
yields two different eigenstates, which correspond to two on-shell
physical mesons and one of them might be identified as $X(5568)$.

Let us designated the two physical states as $S_1$ and $S_2$. We
formulate the mixing matrix as:
\begin{equation}\label{M1}
     \left(\begin{array}{c}
     S_1\\
       S_2
      \end{array}\right)=\left(\begin{array}{cc}
       {\rm sin}\theta &{\rm cos}\theta \\
        -{\rm cos}\theta &   {\rm sin}\theta
      \end{array}\right)\left(\begin{array}{c}
      T\\
       M
      \end{array}\right),
  \end{equation}
where $\theta$ is the assumed mixing angle between the tetraquark
and molecular state.

The masses of the two physical states ($m_{S_1}$ and $m_{S_2}$)
can be obtained as:
\begin{equation}\label{M2}
    \left(\begin{array}{cc}
      m_{S_1} &0 \\
       0 &   m_{S_2}
      \end{array}\right)=\left(\begin{array}{cc}
       {\rm sin}\theta &-{\rm cos}\theta \\
        {\rm cos}\theta &   {\rm sin}\theta
      \end{array}\right)\left(\begin{array}{cc}
      m_T &\Delta \\
      \Delta &   m_M
      \end{array}\right)\left(\begin{array}{cc}
       {\rm sin}\theta &{\rm cos}\theta \\
        -{\rm cos}\theta &   {\rm sin}\theta
      \end{array}\right),
  \end{equation}
where $m_T$ and  $m_M$ are the bare masses of the tetraquark ($T$)
and molecular state ($M$), respectively.

For the evaluation, we need to input $m_T$ and $m_M$. According
to\cite{Feng:2011zzb}, the $BK$ bound state was studied using the
Bethe--Salpeter equation and the binding energy varied from 15 MeV
to 85 MeV, which corresponded to a mass of $5733\pm 35$ MeV. The
molecular state of $BK$ was also estimated with the Quantum
Chromodynamics (QCD) sum rules and the center mass of 5757
MeV\cite{Agaev:2016urs} was obtained with a large uncertainty. In
our calculation, we set $m_M=5733\pm$35 MeV. The mass of the
tetraquark has been estimated in several previous studies and
different values were obtained. Stancu\cite{Stancu:2016sfd}
employed a simple quark model and determined the mass of the
tetraquark as 5530 MeV. In the chiral quark model, Chen et al.
\cite{Chen:2016npt} determined the tetraquark mass as about 6400
MeV. The scalar tetraquark given by\cite{Wang:2016tsi} is 5708
MeV. According to\cite{Burns:2016gvy}, a simple sum of the total
masses of the constituent quarks ($su \bar b\bar d$) could be
close to 6107 MeV (indeed, there is an arbitrariness when
selecting the quark masses). If we set the binding energy as 131
MeV\cite{Wang:2016tsi} or 225 MeV\cite{Liu:2016ogz}, the
tetraquark mass $M_T$ should be 5929$\pm$47 MeV.

After we know the off-diagonal matrix elements, we can diagonalize
the matrix and obtain two eigenstates: $S_1$ and $S_2$, and one of
them should be identified as $X(5568)$, although its mass is not
exactly that determined by the D0 collaboration. We have
\begin{eqnarray}
|S_1> &=& \cos\theta |T>+\sin\theta |M>\\
|S_2> &=& -\sin\theta |T>+\cos\theta |M>,
\end{eqnarray}
where $|T>$ and $|M>$ are the pure tetraquark and molecule state,
respectively.


Now, let us consider the hadronic matrix element of $S_1\to
B_s\pi$. According to\cite{Wang:2016wkj,Dias:2016dme}, $X(5568)\to
B_s\pi$ was calculated in terms of the QCD sum rules while
assuming that $X(5568)$ is a tetraquark. Similar results were
obtained in these two studies. The coupling constant was obtained
as $g_{X_TB_s\pi}=(10.6\pm2.1)$ GeV by\cite{Wang:2016wkj}. We
calculated this transition rate in the light front quark model
while assuming that $X(5568)$ is a molecular
state\cite{Ke:2018jql} and the corresponding coupling constant was
determined as $g_{X_MB_s\pi}=(13.0\pm2.4)$ GeV.

In the mixture scenario, the hadronic transition amplitude is
written as
\begin{equation}
<B_s\pi|H_{eff}|S_1>=\cos\theta<B_s\pi|H_{eff}^{(1)}|T>+\sin\theta<B_s\pi|H_{eff}^{(2)}|M>,
\end{equation}
with $H_{eff}=H_{eff}^{(1)}+H_{eff}^{(2)}$. The effective
interaction $H_{eff}$ can be divided into two parts, where
$H_{eff}^{(1)}$ corresponds to a quark--antiquark exchange between
diquark and antidiquark to make color singlet final mesons,
whereas $H_{eff}^{(2)}$ is responsible for dissolving the
molecular state.

According to our strategy, i.e., letting the contributions of the
tetraquark and molecular state fully cancel each other (almost),
we can fix the mixing angle $\theta$ as $(-50.8\pm7.8)^\circ$ when
$S_1$ is regarded as the narrow $X(5568)$.

After substituting the values of $m_T$, $m_M$ and $\theta$ into
Eq. (\ref{M2}) and diagonalizing the mass matrix, we have two
eigenvalues comprising $m_{S_1}$ and $m_{S_2}$, which are
$m_{S_1}=5344\pm307$ MeV and $m_{S_2}=6318\pm315$ MeV,
respectively. It should be noted that $X(5568)$ lies in the error
range of $m_{S_1}$. In this scheme, another physical state $S_2$
exists that should also decay into $B_s\pi$ because for $S_2$, the
tetraquark and molecule components contribute constructively to
the $B_s+\pi$ final state, and its partial width should be large.
Using the values $g_{X_TB_s\pi}$, $g_{X_MB_s\pi}$ and the mass of
$S_2$, we predict the width as $\Gamma(X_{S_2}\to
B_s\pi)=224\pm97$ MeV. Naturally, the experimental search for a
wide resonance at $B_s\pi^{\pm}$ channels is crucial for testing
our ansatz.

 We can also study the possible charmed partners of these
states. Similar to the $BK$ case, the bare mass of the $DK$
molecular state is $2311\pm35$ MeV, whereas the bare mass of a
tetraquark of $ s u \bar c\bar d$ is $2589\pm 47$ MeV, which is
consistent with that reported by\cite{Agaev:2016lkl}. However, in
our case, the tetraquark of $ s u \bar c\bar d$ is not a physical
state. Under heavy quark symmetry, we use the same mixing angle to
obtain two physical states with masses of $1759\pm414$ MeV and
$3141\pm417$ MeV. By contrast, \cite{Liu:2016ogz} estimated the
mass of the tetraquark state corresponding to X(5568) as 2262 MeV,
which is slightly larger than our estimate for the first charm
partner($1759\pm 414$ MeV).
\section{Conclusion and Discussion}

To reconcile the discrepancy between the results obtained for
$X(5568)$ by the D0 collaboration and most other important
experimental collaborations, we propose that based on the quark
model, an exotic state with four different flavors $(s u\bar b\bar
d)$ should exist but it is a mixture of a tetraquark and hadronic
molecule. There should be two eigenstates comprising $S_1$ and
$S_2$, which are the on-shell physical particles. For the lighter
$S_1$ with a mass similar to that assumed for $X(5568)$, the
tetraquark and molecule components contribute destructively to the
$B_s\pi^{\pm}$ final state. This cancellation allows it to escape
detection. We consider that this cancellation is accidental and
incomplete because no principle can ensure full cancellation.
Therefore, it is possible that one collaboration has observed a
small signal whereas others have not.

In our computations, we employed theoretical estimates of the
masses of the tetraquark and molecular state as inputs. The
hadronic transition matrix elements of
$<B_s\pi^{\pm}|H_{eff}|S_1(S_2)>$ were calculated in different
models, so remarkable theoretical uncertainties might have been
involved. Thus, we cannot guarantee accurate quantitative results,
but the qualitative consequences are reasonable and acceptable.

Moreover, we predicted an extra exotic state $|S_2>$ as the
partner of $|S_1>$ with the same quark contents, but different
combinations of the tetraquark and molecular state. It is
important to test our ansatz by searching for this new particle in
the $B_s\pi^{\pm}$ final state.


\section*{Acknowledgment}
This study was supported by the National Natural Science
Foundation of China (NNSFC) under contract No. 11375128.

\end{document}